\documentclass[twocolumn,floatfix,superscriptaddress,longbibliography]{revtex4-1}

\usepackage{amssymb,amsmath,bm}
\usepackage{graphicx}
\usepackage{epstopdf}
\usepackage{color}
\usepackage{appendix}
\usepackage[T1]{fontenc}
\usepackage[colorlinks=true,citecolor=blue,linkcolor=magenta]{hyperref}

\newcommand{\bbA}{\mathbb{A}}

\newcommand{\bbI}{\mathbb{I}}
\newcommand{\bbF}{\mathbb{F}}
\newcommand{\bbM}{\mathbb{M}}

\newcommand{\bbX}{\mathbb{X}}

\newcommand{\mA}{\mathcal{A}}
\newcommand{\mB}{\mathcal{B}}
\newcommand{\mC}{\mathcal{C}}

\newcommand{\ket}[1]{\mbox{$| #1 \rangle$}}

\begin{document}

\title{Determining non-Abelian topological order from infinite projected entangled pair states}

\author{Anna Francuz}
\email[corresponding author: ]{anna.francuz@uj.edu.pl}
\affiliation{Institute of Theoretical Physics, Jagiellonian University, 
             ul. {\L}ojasiewicza 11, PL-30-348 Krak\'ow, Poland}

\author{Jacek Dziarmaga}
\affiliation{Institute of Theoretical Physics, Jagiellonian University, 
             ul. {\L}ojasiewicza 11, PL-30-348 Krak\'ow, Poland}

\date{\today}

\begin{abstract}
We generalize the method introduced in Phys. Rev. B 101, 041108 (2020) of extracting information about topological order from the ground state of a strongly correlated two-dimensional system represented by an infinite projected entangled pair state (iPEPS) to non-Abelian topological order. When wrapped on a torus the unique iPEPS becomes a superposition of degenerate and locally indistinguishable ground states. 
We find numerically symmetries of the iPEPS, represented by infinite matrix product operators (MPO), and their fusion rules. The rules tell us how to combine the symmetries into projectors onto states with well defined anyon flux. 
A linear structure of the MPO projectors allows for efficient determination for each state its second Renyi topological entanglement entropy on an infinitely long cylinder directly in the limit of infinite cylinder's width. 
The same projectors are used to compute topological $S$ and $T$ matrices encoding mutual- and self-statistics of emergent anyons.
The algorithm is illustrated by examples of Fibonacci and Ising non-Abelian string net models.
\end{abstract}

\maketitle


\section{Introduction}

Topologically ordered phases \cite{wen1990topological} support anyonic quasiparticles. They open the possibility of realizing fault-tolerant quantum computation \cite{kitaev2003fault-tolerant} based on braiding of non-Abelian anyons. Apart from a number of exactly solvable models \cite{kitaev2003fault-tolerant, kitaev2006anyons, levin2005string-net}, verifying whether a given microscopic Hamiltonian realizes a topologically ordered phase has traditionally been regarded as an extremely hard task. Recently, observation of quantized Hall effect in Kitaev-like ruthenium chloride $\alpha$-$RuCl_3$ in magnetic field \cite{KitaevExp} granted the problem with urgent experimental relevance. 


\begin{figure}[t!]
\includegraphics[width=\columnwidth]{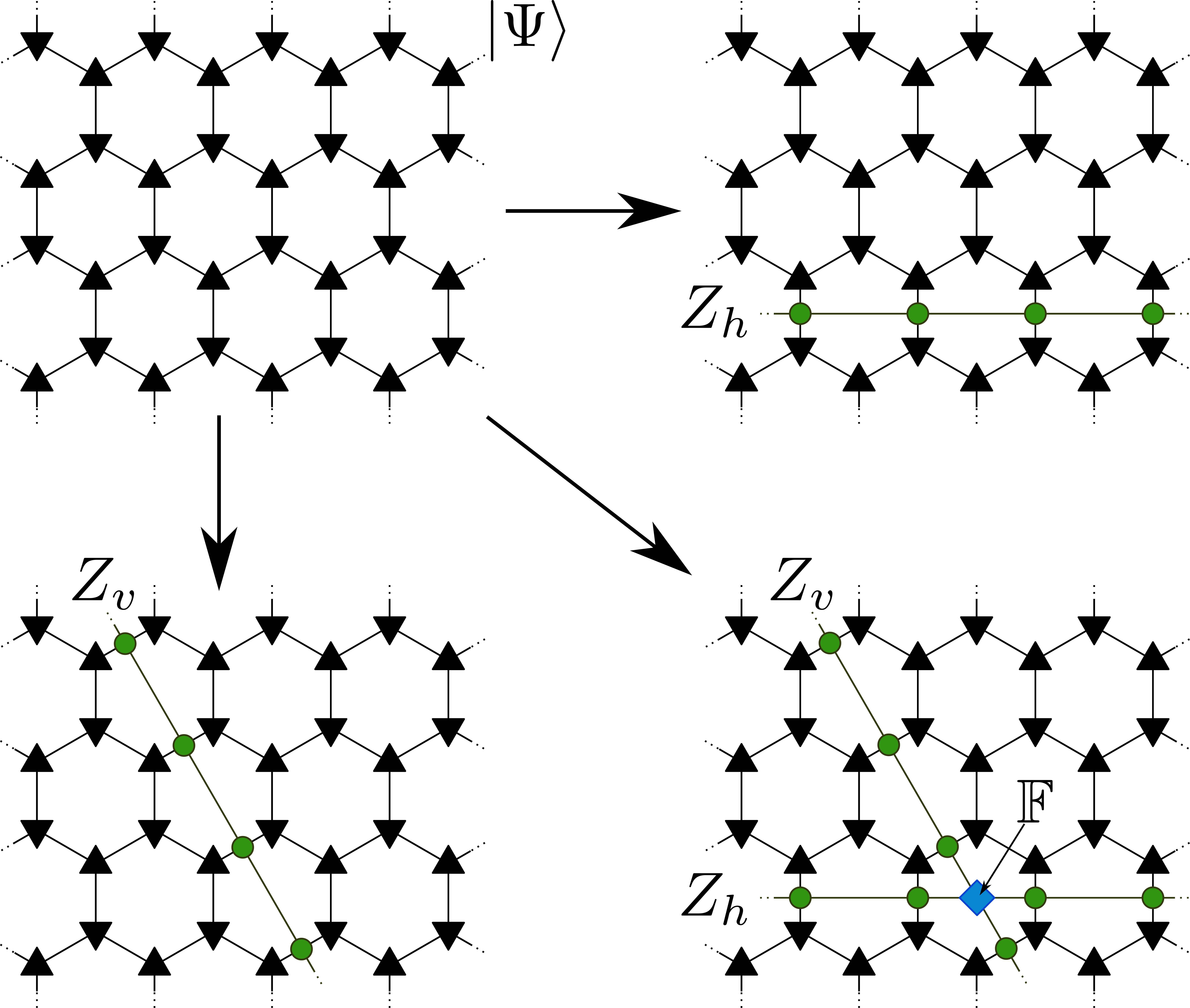}
\caption{ 
{\bf General picture.} 
From the unique ground state on an infinite lattice represented by an iPEPS $\ket{\Psi}$, we construct various states inserted with MPO symmetries. Their linear combinations, whose coefficient are determined by fusion rules of the MPO symmetries $Z_{h,v}$ (corresponding to anyonic fusion rules), become a basis of states with well defined anyonic flux.
Here physical indices are not drawn for simplicity.
}
\label{fig:summary}
\end{figure}


A leading numerical method is to use density matrix renormalization group (DMRG) \cite{white1992density, white1993density} on a long cylinder \cite{yan2011spinliquid, jiang2012identifying, gong2013phase, zhu2013weak, gong2014emergent, zhu2014quantum, gong15global, hu2015topological, zhu15emergence, zhu2015spin, zaletel2016space, zeng2017nature, vaezi2017numerical, zhu2018robust, PollmannKitaevlike, PollmannKitaev}. In the limit of infinitely long cylinders, DMRG naturally produces ground states with well-defined anyonic flux, from which one can obtain full characterization of a topological order, via so-called topological $S$ and $T$ matrices \cite{cincio2013characterizing}. Since the proposal of Ref.~[\onlinecite{cincio2013characterizing}], this approach has become a common practice \cite{he2014chiral, zhu2014topological, zhu2014chiral, bauer2014chiral, zhu2015fractional, grushin15characterization, he2015kagome, he15distinct, he15bosonic,  geraedts15competing, mong2015fibonacci, he17realizing, stoudenmire2015assembling, he17signatures, Saadatmand20016symmetry, hickey2016haldane, zaletel2017measuring, zeng2018tunning}.

Unfortunately, the cost of a DMRG simulation grows exponentially with the circumference of cylinder, limiting this approach to thin cylinders (up to a width of $\simeq 14$ sites) and short correlation lengths (up to $1-2$ sites). Instead, infinite projected entangled pair states (iPEPS) in principle allow for much longer correlation lengths \cite{verstraete2004renormalization, murg2007variational, verstraete2008matrix}. A unique ground state on an infinite lattice can be represented by an iPEPS that is either a variational ansatz \cite{chiralpeps} or a result of numerical optimization \cite{varCorboz,topoAF1}. When wrapped on a cylinder the iPEPS becomes a superposition of degenerate ground states with definite anyonic fluxes. Here we generalize the approach of Ref. \cite{topoAF1} to non-Abelian topological order and show how to produce a PEPS-like tensor network for each ground state with well-defined flux. Such tensor networks are suitable for extracting topological $S$ and $T$ matrices by computing overlaps between ground states. Furthermore, we show that they allow for computation of topological second Renyi entropy directly in the limit of infinite cylinder's width. The approach of Ref. \cite{topoAF1} does not assume clean realization of certain symmetries on the bond indices, in contrast to \cite{burak2014characterizing, bultinck2017anyons, iqbal2018study, fernandez2016constructing}. This has been demonstrated in Ref. \cite{topoAF1} by examples of toric code and double semions perturbed away from a fixed point towards a ferromagnetic phase as well as for the numerical iPEPS representing the ground state of the Kitaev model in the gapped phase. The last example shows that the method does not require restoring the symmetries by suitable gauge transformations of a numerical iPEPS, a feat that was accomplished in Ref. \cite{topoCorboz} for the toric code with a perturbation. Finally, it also has much lower cost than methods based on the tensor renormalization group \cite{he2014modular}.

The ferromagnetic Kitaev model in a weak $(1,1,1)$ magnetic field supports non-Abelian chiral topological order \cite{kitaev2006anyons,PollmannKitaev} and Ref. \cite{KitaevExp} is believed to provide the first experimental realization of this universality class. However, as the magnetic field is a tiny perturbation of a critical state, the correlation length should be long \cite{chiralpeps}. This drives the problem beyond accurate DMRG simulation on a thin cyllinder and, therefore, the non-Abelian phase observed in the experiment \cite{KitaevExp}, may require iPEPS for its accurate description.

In this work we consider mainly string-net models. The key elements of the method introduced in Ref. \cite{topoAF1} are shown in Fig. \ref{fig:summary}. Virtual indices of iPEPS on a torus or cylinder can be inserted with horizontal/vertical matrix product operator (MPO) symmetries. Their action on iPEPS is the same as flux operators (~Wilson loops) winding around the torus in the same horizontal/vertical direction. However, the MPO symmetries are much easier to find than the non-local operators that -- in interacting systems -- become complicated operator ribbons rather than simple strings. Just as projectors on definite anyon fluxes could be in principle constructed as linear combinations of flux operators, virtual projectors can be made as combinations of the MPO symmetries.  

The paper is organized in sections \ref{sec:MPOsymmetries}...\ref{sec:STmatrices} where we gradually introduce subsequent elements of the algorithm. Most sections open with a general part introducing a new concept. Then a series of subsections follows illustrating the general concept with a series of examples: the Abelian toric code (to make contact with Ref. \cite{topoAF1}), Fibonacci string net, and Ising string net. In the end the algorithm is summarized in section \ref{sec:summary}. Additionally, in appendix \ref{sec:kitaev} we apply some of the same tools to a variational ansatz proposed for the Kitaev model in magnetic field \cite{chiralpeps}. A detailed plan is as follows.

In Sec. \ref{sec:MPOsymmetries} we define fixed points of the iPEPS transfer matrix in the form of MPS and introduce MPO symmetries that map between different fixed points. We also identify fusion rules of the MPO symmetries that are isomorfic with anyonic fusion rules.
In Sec. \ref{sec:verticalprojectors} we consider an iPEPS wrapped on an infinite cylinder -- that we visualize as horizontal without loss of generality -- and use the fusion rules to construct vertical projectors on states with definite anyon flux along the horizontal cylinder. 
In Sec. \ref{sec:IMPOsymmetries} we consider again an iPEPS wrapped on an infinite cylinder but this time the iPEPS is inserted with a horizontal MPO symmetry that alters boundary conditions in the vertical direction. We construct its vertical MPO symmetries that we call impurity MPO (IMPO) symmetries. We also identify their fusion rules.
In Sec. \ref{sec:impurityprojectors} the fusion rules are used to construct vertical projectors as linear combinations of the IMPO symmetries. The impurity projectors select states with definite horizontal anyon flux in the iPEPS inserted with the horizontal MPO symmetry.
In Sec. \ref{sec:TEEverticalprojectors} we show how the structure of vertical projectors enables efficient evaluation of the topological second Renyi entanglement entropy directly in the limit of infinite cylinder's width.
In Sec. \ref{sec:TEEimpurityprojectors} the same is done with impurity projectors.
Finally, in Sec. \ref{sec:STmatrices} we show how to obtain the topological $S$ and $T$ matrices from overlaps between states with definite anyon flux. In case of string net models they provide full characterization of the topological order.
The paper is closed with a brief summary in section \ref{sec:summary}.


\begin{figure}[t!]
\includegraphics[width=\columnwidth]{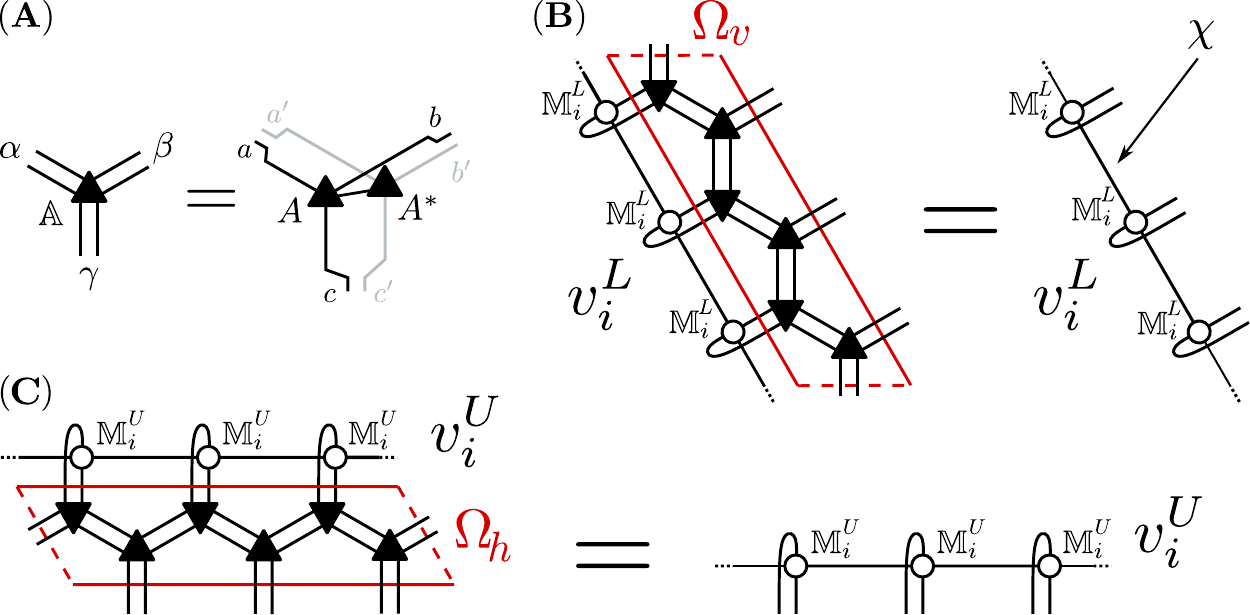}
\caption{
{\bf Transfer matrix.}
In (A), 
graphical representation of a double tensor $\bbA$.
In (B), 
leading left eigenvector $\left(v_i^L\right|$ of vertical transfer matrix $\Omega_v$ takes an MPO form $v_i^L$. The uniform $v_i^L$ is made of tensors $\bbM_i^{L}$ with bond dimension $\chi$ that can be obtained with the VUMPS algorithm \cite{VUMPS2,VUMPS}.
In (C), 
up eigenvector $v_i^U$ of horizontal TM $\Omega_h$.
}
\label{fig:TM}
\end{figure}


\section{Generators of symmetries}
\label{sec:MPOsymmetries}

Uniform iPEPS on a honeycomb lattice can be characterized by a tensor $A$ with elements $A^i_{abc}$. Here, $i$ is a physical index and $a,b,c$ are bond indices. Let $\bbA$ denote a double tensor $\bbA = \sum_i A^i \otimes (A^i)^\ast$ with double bond indices $\alpha = (a,a')$, etc., see Fig.~\ref{fig:TM}(A) and appendix \ref{app:tensors}. iPEPS transfer matrix (TM) $\Omega$ is defined by a line of double tensors $\bbA$ contracted via their bond indices along the line as shown in Figs.~\ref{fig:TM}(B) and (C). These figures show vertical TM $\Omega^v$ and horizontal TM $\Omega^h$, respectively. Their leading eigenvectors are TM fixed points. In the thermodynamic limit only the leading eigenvectors survive in TM's spectral decomposition: 
\begin{eqnarray}
\Omega^v 
   \approx 
   \omega 
   \sum_{i=1}^n \left.\vert v_i^R\right)\left(v_i^L\vert\right.,~~
\Omega^h
   \approx 
   \omega 
   \sum_{i=1}^n \left.\vert v_i^U\right)\left(v_i^D\vert\right..
\end{eqnarray}
The leading eigenvalue, $\omega$, is the same for both vertical and horizontal TM. The leading eigenvectors are biorthonormal: 
\begin{eqnarray}
&& \delta_{ij}=\left(v_i^{L}\vert v_j^{R}\right)={\rm Tr}~ \left( v_i^L \right)^T v_j^R ,\\
&& \delta_{ij}=\left(v_i^{U}\vert v_j^{D}\right)={\rm Tr}~ \left( v_i^U \right)^T v_j^D,
\end{eqnarray}
Here we use both the MPS, $\left|v_i\right)$, and MPO, $v_i$, forms. MPS $\left|v_i\right)$ is MPO $v_i$ between bra and ket indices of the double iPEPS TM. The ansatz for a fixed point boundary $v_i^X$ is a pure MPO with spectral radius $1$ \cite{IrreducibleMPS} made out of tensors $M_i^X$. 

Different fixed points are connected by symmetries whose existence is a distinctive feature of topologically ordered states encoded in iPEPS. In contrast, in the trivial ferromagnetic phase the two boundary fixed points, $v_\uparrow$ and $v_\downarrow$, corresponding to two different magnetizations have orthogonal support spaces and, therefore, the operator mapping between them does not exist. The symmetries act on virtual indices of the tensor network. They are called MPO symmetries and, apart from few exactly solvable models for which they can be found analytically \cite{burak2014characterizing}, they have to be found numerically as described in \cite{topoAF1}. The MPO symmetries $Z_a$ are operators which form certain algebra under their multiplication:
\begin{equation}
Z_a Z_b = \sum_c N^c_{ab}~ Z_c,
\label{ZN}
\end{equation}
where the possible values of $N^c_{ab}$ are $0,1$. Each MPO symmetry $Z_a$ (including the trivial identity $Z_1\equiv 1$) corresponds to certain anyon type $a$ in a sense that their algebra is the same as the fusion rules of the anyons, see appendix \ref{app:FR}. Once all boundary fixed points $v_i$ are found numerically, the MPO symmetries $z_{ij}$ are obtained as MPO's mapping between the boundaries: 
\begin{equation}
v_i\cdot z_{ij} = v_j.
\end{equation} 
The same set of symmetries exists for $L/R$ and $U/D$ boundary fixed points. We completed these numerical procedures in the following models.

\subsection{Toric code}

We begin with this basic example to make contact with Ref. \cite{topoAF1} where the Abelian version of the present method was applied to this model and its realistic implementation with Kitaev model \cite{kitaev2006anyons}. Each TM has $2$ boundary fixed points. To be more specific, for vertical transfer matrix $\Omega^v$ in addition to $Z^v_1=1$ we find numerically one non-trivial MPO-symmetry $z_{12}^v=z_{21}^v\equiv Z_2^v$ that satisfies
\begin{equation}
    v_1^L\cdot Z_2^v=v_2^L,~~v_2^L\cdot Z_2^v=v_1^L.
    \label{vz-toric}
\end{equation}
These equations imply ${\cal Z}_2$ algebra:
\begin{equation}
    Z_2^v \cdot Z_2^v = 1.
    \label{toricFR}
\end{equation}
It has to be strongly emphasized that in general the numerical solution $Z_2^v$ of equations (\ref{vz-toric}) has zero modes that make the algebra valid only in the sense that $v_i^L\cdot Z_2\cdot Z_2=v_i^L$ for any $i$. The same reservation applies to all fusion rules (\ref{ZN}) to be identified numerically in the rest of this paper. This is also why all (numerically obtained) MPO symmetries throughout the paper are used only in iPEPS embedding: the zero modes do not matter when inserted between columns/rows of an iPEPS. Keeping this in mind, for all fixed point tensors considered in this paper the algebra (\ref{ZN}) is satisfied with close to machine precision.

\subsection{Fibonacci string-net}

Here we employed the iPEPS tensors for a fixed point Fibonacci string net model presented in appendix \ref{app:tensors}. For each TM we found numerically $2$ boundary fixed points and one non-trivial MPO symmetry $Z_2$ satisfying, e. g.,
\begin{equation}
    v_1^L\cdot Z_2^v=v_2^L.
\end{equation}
The same MPO was found to satisfy also
\begin{equation}
    v_2^L\cdot Z_2^v=v_1^L+v_2^L.
\end{equation}
These two equations imply the Fibonacci fusion rule
\begin{equation}
    Z_2^v\cdot Z_2^v=1^v+Z_2^v.
    \label{FibonacciFR}
\end{equation}
Again, due to zero modes, the rule holds only when applied to iPEPS boundaries. Similar MPO symmetries were also found for the horizontal boundary fixed points.

\subsection{Ising string net}

Here we employed the iPEPS tensors for a fixed point Ising string net model presented in appendix \ref{app:tensors}. This time each TM has $3$ boundary fixed points. We found two non-trivial MPO symmetries, labelled as $Z_\sigma$ and $Z_\psi$, as numerical solutions to equations, e. g., 
\begin{equation}
    v_1^L\cdot Z_\sigma^v=v_2^L,~~v_1^L\cdot Z_\psi^v=v_3^L.
\end{equation}
Furthermore, we found that the solutions satisfy
\begin{equation}
    v_2^L\cdot Z_\sigma^v=v_1^L+v_3^L,~
    v_3^L\cdot Z_\psi^v=v_1^L,~
    v_2^L\cdot Z_\psi^v=v_2^L,~
    v_3^L\cdot Z_\sigma^v=v_2^L.
\end{equation}
These six equations imply non-trivial fusion rules:
\begin{equation}
    Z_\sigma^v \cdot Z_\sigma^v = 1^v + Z_\psi^v ,~ 
    Z_\sigma^v \cdot Z_\psi^v = Z_\sigma^v = Z_\psi^v\cdot Z_\sigma^v ,~
    Z_\psi^v\cdot Z_\psi^v=1^v.
    \label{IsingFR}
\end{equation}
which justify the labelling. For our numerical $Z_\psi^v$ and $Z_\sigma^v$ the rules hold only when applied to $v_i^L$. Similar MPO symmetries were also found for the horizontal boundary fixed points.


\section{Vertical projectors}
\label{sec:verticalprojectors}

The MPO symmetries alone are enough to construct some of the projectors on states with definite anyon fluxes. Let us consider vertical MPO symmetries $Z^v_a$ for definiteness. Their linear combinations 
\begin{equation}
    P=\sum_a c_a Z^v_a,
    \label{Pvertical}
\end{equation}
which satisfy $P\cdot P=P$, make vertical projectors. When these projectors are inserted into iPEPS wrapped on an infinite horizontal cylinder, they yield states with definite anyon fluxes along that cylinder. The remaining projectors that can be applied when the iPEPS is inserted with a line of $Z^h$ are subject of the following section.

\subsection{Toric code}

The ${\cal Z}_2$ algebra (\ref{toricFR}) allows for two projectors,
\begin{equation}
    P_{\pm}=\frac12\left(1\pm Z^v_2\right),
\end{equation}
that satisfy $P_\pm\cdot P_\pm=P_\pm$ and $P_+\cdot P_-=0=P_-\cdot P_+$. Later on they will be identified as $P_+\equiv P_{\rm vac}$ and $P_-\equiv P_e$, i.e., projectors on the vacuum and the electric flux, respectively.

\subsection{Fibonacci string net}

The fusion rules (\ref{FibonacciFR}) determine two projectors:
\begin{equation}
    P_\pm=\frac{1}{\sqrt5}\left(\phi^{\pm1} 1 \mp Z^v_2\right).
\end{equation}
Here $\phi=(\sqrt5+1)/2$. They will be identified as $P_+\equiv P_{\rm vac}$ and $P_-\equiv P_{\tau\bar\tau}$, i.e., projectors on the vacuum and the sector with both Fibonacci anyons: $\tau$ and $\bar\tau$.

\subsection{Ising string net}

The fusion rules (\ref{IsingFR}) allow for six projectors:
\begin{eqnarray}
P_{1,2}&=&\frac12\left(1^v\pm Z_\psi^v\right),\\
P_{3,4}&=&\frac14\left(3~1^v-Z_\psi^v\right) \pm \frac{1}{\sqrt8}Z_\sigma^v,\\
P_{5,6}&=&\frac14\left( 1^v+Z_\psi^v\right) \pm \frac{1}{\sqrt8}Z_\sigma^v.
\end{eqnarray}
Not all of them are the minimal projectors on definite anyon flux. It is easy to check that $P_3\cdot P_4=P_2$ and, therefore, out of the three it is enough to keep only $P_2$. Furthermore, we can see that $P_5+P_6=P_1$ hence we can skip $P_1$. After this selection we are left with three minimal projectors $P_{2,5,6}$ that satisfy $P_a\cdot P_b=P_a\delta_{ab}$. They will be identified as $P_5\equiv P_{\rm vac}$, $P_6\equiv P_{\psi\bar\psi}$, and $P_2\equiv P_{\sigma\bar\sigma}$.


\begin{figure}[t!]
\includegraphics[width=\columnwidth]{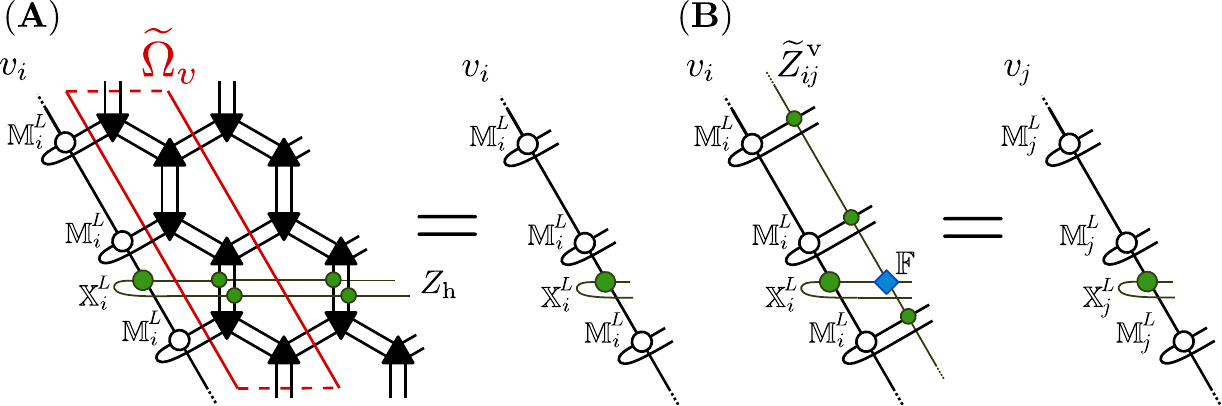}
\caption{
{\bf Impurity transfer matrix.}
In (A),
with $Z^h$ inserted into both bra and ket layers of the iPEPS the transfer matrix $\Omega^v$ becomes impurity transfer matrix $\tilde\Omega^v$.
Its leading left eigenvectors $\left(x^L\right|$ are obtained from MPOs from $v^L$ by inserting additional tensors $\bbX^L$. Here double lines are dropped to improve clarity.
In (B), 
graphical illustration of Eq.~(\ref{xFx}).
}
\label{fig:impurityTM}
\end{figure}


\section{Impurity MPO symmetries}
\label{sec:IMPOsymmetries}

In order to construct the remaining projectors, that are to be applied to an iPEPS inserted with a nontrivial horizontal MPO symmetry $Z^h$, we need to introduce an impurity transfer matrix (ITM), see Fig. \ref{fig:impurityTM} (A). In general ITM has a number of leading left and right eigenvectors, respectively $\left(x^L_i\right|$ and $\left|x^R_j\right)$, that are biorthonormal: $\left(x^L_i|x^R_j\right)=\delta_{ij}$. The eigenvectors are constructed by inserting the eigenvectors of the vertical TM, respectively $v^L$ and $v^R$, with additional tensors $\bbX^L_i$ and $\bbX^R_j$, see Fig. \ref{fig:impurityTM} (A). The same figure shows equations that need to be satisfied by the additional tensors. They are efficiently obtained from a generalized eigenvalue problem:
\begin{equation}
    \left(x^L_i\right|\tilde{\Omega}_v\left|x^R_j\right) =\lambda \left(x^L_i|x^R_j\right).
\end{equation}
Here $\lambda=1$ is the maximal generalized eigenvalue. The problem is to be understood as
\begin{equation}
    \left(\bbX^L_i\right)^T \cdot {\cal M} \cdot \bbX^R_j = 
    \lambda \left(\bbX^L_i\right)^T \cdot {\cal N} \cdot \bbX^R_j,
\end{equation}
where $\bbX^L_i$ and $\bbX^R_j$ are vectorized and matrices ${\cal M}$ and ${\cal N}$ are tensor environments of $\bbX^L_i$ and $\bbX^R_j$ in $\left(x^L_i\right|\tilde{\Omega}_v\left|x^R_j\right)$ and $\left(x^L_i|x^R_j\right)$, respectively.

Furthermore, as shown in Fig. \ref{fig:impurityTM} (B), left eigenvector $\left(x^L_i\right|$ can be acted on by any vertical MPO symmetry $Z^v$, including the trivial identity $Z^v_1=1^v$. In order to make the action possible, $Z^v$ has to be inserted with additional tensor $\bbF$ that acts on $Z^h$. With appropriate choice of $\bbF_{ij}$ their combination gives rise to impurity MPO-symmetry $\widetilde z^v_{ij}$ such that
\begin{equation}
    x^L_i \widetilde z_{ij} = x^L_j.
    \label{xFx}
\end{equation}
A necessary condition for symmetry $\widetilde z^v_{ij}$ to exist is that $v^L$ in $x^L_i$, here denoted by $v^L(i)$, and $v^L$ in $x^L_j$, here denoted by $v^L(j)$, are related by $v^L(i)\cdot Z^v=v^L(j)$.

A straightforward but essential observation is that, in analogy to MPO symmetries, the IMPO symmetries also satisfy their own fusion rules:
\begin{equation}
    \widetilde Z^v_a \cdot \widetilde Z^v_b =
    \sum_c \widetilde N^{c}_{ab} \widetilde Z^v_{c}.
    \label{impurityZFR}
\end{equation}
Here we keep only the minimal set of independent IMPO symmetries denoted by a capital $\widetilde Z$ and labelled with a single index $a,b,c$. In general the coefficients $\widetilde N^{c}_{ab}$ do not need to be integers as they depend on normalization of the eigenevectors $\left(x^L_i\right|$ and $\left|x^R_j\right)$. 


\section{Impurity projectors}
\label{sec:impurityprojectors}

In analogy to the vertical MPO symmetries and vertical projectors, as a product of two IMPO symmetries is a linear combination of IMPO symmetries, see Eq. (\ref{impurityZFR}), we can find projectors as linear combinations of IMPO symmetries,
\begin{equation}
    \widetilde P = \sum_{a} \tilde c_{a} \widetilde Z^v_{a}.
\end{equation}
The condition $\widetilde P\cdot \widetilde P=\widetilde P$ is equivalent to a set of quadratic equations for coefficients $\tilde c_{a}$. Numerically it seems more efficient to find the coefficients by repeated Lanczos iterations: 
\begin{equation}
    \widetilde P' \propto \widetilde P\cdot\widetilde P. 
    \label{Lanczos}
\end{equation}
In each iteration the IMPO fusion rules (\ref{impurityZFR}) are used to express the product $\widetilde P\cdot\widetilde P$ as a new linear combination $\widetilde P' = \sum_{a} \tilde c'_{a} \widetilde Z^v_{a}$
and then new coefficients $\tilde c'_{a}$ are normalized so that the maximal magnitude of the eigenvalues of $\widetilde P'$ is $1$. Therefore, each iteration is a map $\{c_a\}\rightarrow\{c_a'\}$ which is repeated until the coefficients converge. These computations are performed in the biorthonormal eigenbasis of impurity eigenevectors, $\left( x^L_a \right|$ and $\left| x^R_a \right)$, where all involved MPO's become small matrices like, e.g., $\left( x^L_a \right|\widetilde Z^v_c\left| x^R_b\right)\equiv\left[Z^v_c\right]_{ab}$. Repeating the Lanczos scheme with random initial coefficients we obtain all impurity projectors.

\subsection{Toric code}

There is one ITM with $Z^h=Z_2^h$. It has two eigenvectors $\left(x^L_a\right|$, one for each TM eigenvector $v^L_a$. In addition to an identity, $\widetilde 1^v$, there is one non-trivial IMPO symmetry $\widetilde z^v_{12}=\widetilde z^v_{21}\equiv \widetilde Z^v_2$. A non-trivial fusion ${\cal Z}_2$ algebra, $\widetilde Z^v \cdot \widetilde Z^v=1^v$, implies two projectors:
\begin{equation}
    \widetilde P_{\pm}=\frac12\left(\widetilde 1^v\pm \widetilde Z^v_2\right).
\end{equation}
They will be identified as magnetic and fermionic projectors, $\widetilde P_+\equiv\widetilde P_{\rm m}$ and $\widetilde P_-\equiv\widetilde P_\epsilon$, respectively.

\subsection{Fibonacci string net}

There is one ITM with $Z^h=Z^h_2$. It has one eigenvector $\left(x^L_1\right|$ embedded in $v_1^L$ and two eigenvectors $\left(x^L_{2,3}\right|$ embedded in $v_2^L$. We choose the two to be Hermitean and orthonormal but this still leaves (gauge) freedom of their rotation. In addition to the trivial identity, $\tilde 1^v$, there are two ITM symmetries: $\widetilde z^v_{12}$ and $\widetilde z^v_{13}$. Their fusion rules do depend on the gauge but independently of the gauge we find numerically three projectors $\widetilde P_{1,2,3}$. Only two of them project on states that are orthogonal to the states obtained with vertical projectors, as can be verified by calculating overlaps between their respective projected iPEPS on infinite torus. The new projectors will be identified as $\widetilde P_1\equiv \widetilde P_\tau$ and $\widetilde P_2\equiv \widetilde P_{\bar\tau}$. 

Interestingly, the third one, $\widetilde P_3$, projects on the same horizontal anyon flux as vertical projector $P_-$ and both will be identified as $\widetilde P_{\tau\bar\tau}$ and $P_{\tau\bar\tau}$, respectively. This way we have two equivalent ways to obtain ${\tau\bar\tau}$ flux: one with and one without $Z^h_2$ MPO symmetry. In other words, with or without inserted $Z^h_2$ symmetry the iPEPS wrapped on an infinite cylinder has a non-zero overlap with the ground state with ${\tau\bar\tau}$ flux.

\subsection{Ising string net}

There are two ITM with $Z_\sigma^h$ and $Z_\psi^h$. For each of them independently we construct impurity projectors. In case of $Z_\sigma^h$ we find four projectors to be identified later as $\widetilde P_\sigma$, $\widetilde P_{\bar\sigma}$, $\widetilde P_{\sigma\bar\psi}$, and $\widetilde P_{\psi\bar\sigma}$. In case of $Z_\psi^h$ we find three projectors to be identified as $\widetilde P_\psi$, $\widetilde P_{\bar\psi}$, and $\widetilde P_{\sigma\bar\sigma}$. The last one provides a new way to obtain $\sigma\bar\sigma$ flux in addition to vertical projector $P_2\equiv P_{\sigma\bar\sigma}$. This is similar redundancy as in the Fibonacci model.  


\begin{figure}[t!]
\includegraphics[width=\columnwidth]{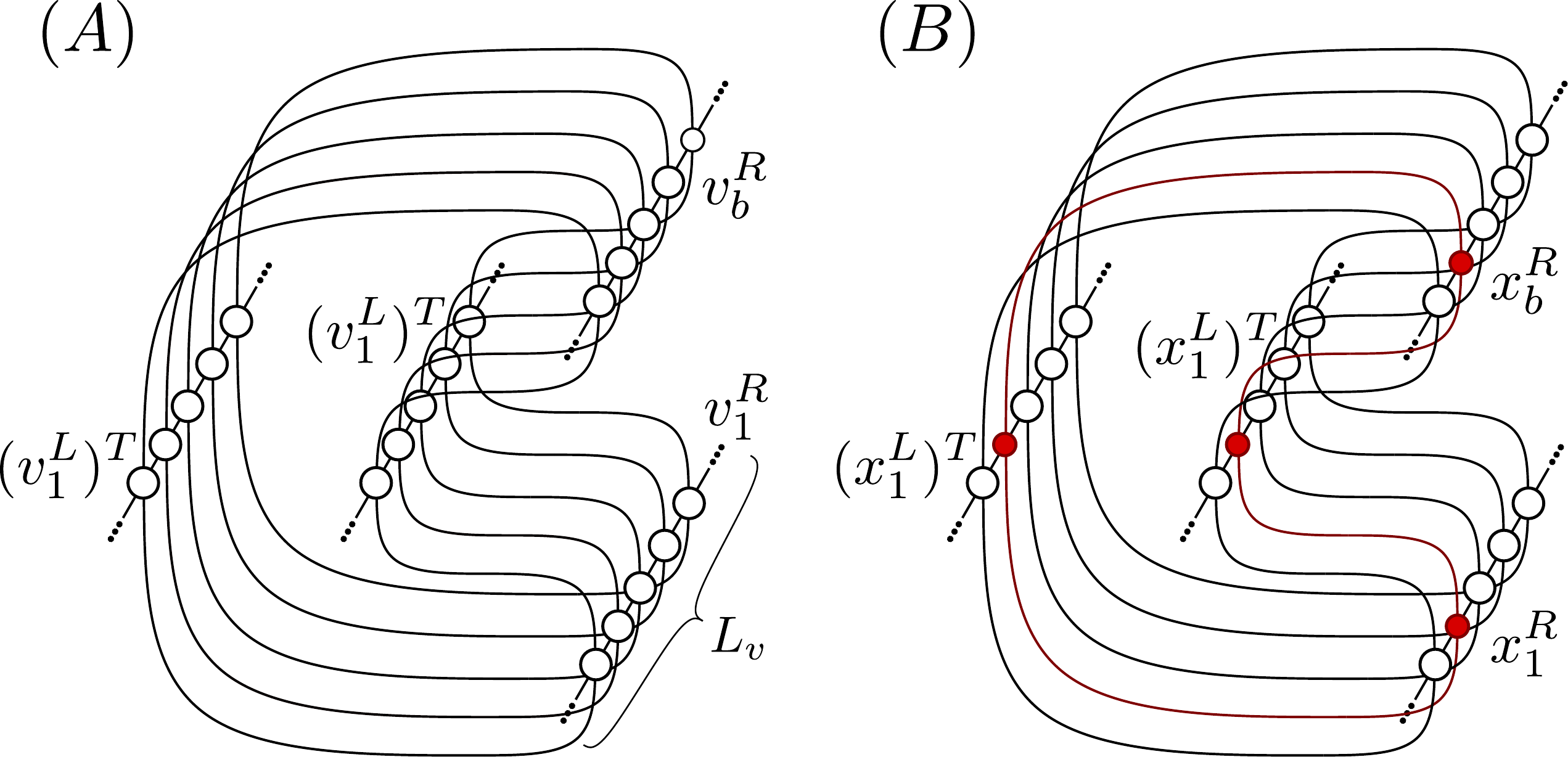}
\caption{
{\bf Topological entropy.}
In (A),
tensor network representing 
$\left( v_{1}^L \right)^T v_{1}^R \left( v_{1}^L\right)^T v_{b}^R$ 
on a vertical cut in an infinite horizontal cylinder of vertical width $L_v$. 
The network is $L_v$-th power of a transfer matrix.
In (B),
tensor network representing 
$\left( x_1^L \right)^T x_1^R \left( x_1^L \right)^T x_b^R$. 
The network is $L_v$-th power of the same transfer matrix inserted with a layer of impurities $\bbX^{L,R}_b$.
}
\label{fig:entropyV}
\end{figure}


\section{ Topological entropy: vertical projectors }
\label{sec:TEEverticalprojectors}

The topological entanglement entropy (TEE) \cite{TopoEntropyKitaevPreskill} is not full characterization of topological order but it may provide quick and numerically stable diagnostic for an iPEPS obtained by numerical minimization. Studies of von Neumann TEE of PEPS wavefunctions have long tradition \cite{cirac2011entanglement} but they require finding full entanglement spectrum of an infinite half-cylinder and extrapolation to the limit of its infinite width, a task that may be hard to accomplish for a long correlation length. In contrast, the projector formalism is naturally compatible with the second Renyi entropy allowing for its efficient evaluation directly in the thermodynamic limit. What is more, in the realm of string net models the Renyi and von Neumann TEE were shown to be the same \cite{RenyiWen}. 

Here we consider a vertical cut in an iPEPS wrapped on an infinite horizontal cylinder of width $L_v$. Its right/left boundary fixed point on the left/right half-cylinder is $\sigma_L$/$\sigma_R$. A reduced density matrix for a half cylinder is isomorfic to \cite{cirac2011entanglement} 
\begin{equation}
    \rho \propto \sqrt{\sigma_L^T} \sigma_R \sqrt{\sigma_L^T}
    \label{rhoL}
\end{equation} 
and its second Renyi entropy is 
\begin{equation}
    S_2 = -\log {\rm Tr}~ \rho^2 = -\log {\rm Tr}~ \sigma_L^T\sigma_R\sigma_L^T\sigma_R.
\end{equation} 
We want the entropy in a state with a definite anyon flux $a$ along the cylinder. 

Towards this end, we begin with $\sigma_{L,R}\propto v_1^{L,R}$ that is a combination of all anyon fluxes.
After inserting projector $P_a$ into the vertical cut we obtain
\begin{equation}
   \rho_a=
   {\cal N}_a
   \sqrt{\left(v_1^L\right)^T} \left( v^R_1 \cdot P_a^T \right) \sqrt{\left(v_1^L\right)^T}.    
\end{equation}
Here the projector was applied to $\sigma_R\propto v^R_1$ without loss of generality and normalization ${\cal N}_a$ is such that ${\rm Tr} \rho_a=1$. The entropy becomes
\begin{eqnarray}
    S_2(a) &=& 
    -\log {\rm Tr}~ \rho^2_a \nonumber\\
           &=& 
    -\log {\cal N}_a^2 ~ {\rm Tr}~ 
                    \left(v_1^L\right)^T \left( v^R_1 \cdot P_a^T \right)
                    \left(v_1^L\right)^T \left( v^R_1 \cdot P_a^T  \right) \nonumber\\
           &=& 
    -\log {\cal N}_a^2 ~ {\rm Tr}~ 
                    \left(v_1^L\right)^T v^R_1
                    \left(v_1^L\right)^T P_a^* v^R_1 \cdot P_a^T \nonumber\\ 
           &=&                     
    -\log {\cal N}_a^2 ~ {\rm Tr}~ 
                    \left(v_1^L\right)^T v^R_1
                    \left(v_1^L\right)^T v^R_1 P_a^T P_a^T \nonumber\\                    
           &=& 
    -\log {\cal N}_a^2 ~ {\rm Tr}~
                    \left(v_1^L\right)^T v^R_1
                    \left(v_1^L\right)^T \left( v^R_1 \cdot P_a^T \right).                     
\end{eqnarray} 
Here we used
$
P_a^T \left(v_1^L\right)^T = \left(v_1^L\right)^T P_a^*
$
and 
$
P_a^* v_1^R = v_1^R P_a^T
$
that follow from the fact that $v$'s are edges of a double-layer iPEPS with bra and ket layers. In this way we are left with only one projector that yields a linear combination,
\begin{equation}
    v^R_1 \cdot P_a^T = \sum_b s^a_b v^R_b,
\end{equation} 
with coefficients $s^a_b$ that follow from the properties of the MPO symmetries whose linear combination is $P_a$. 

The normalization ${\rm Tr}~\rho_a=1$ and the biorthonormality, ${\rm Tr} \left(v_1^L\right)^T v^R_b=\delta_{1b}$, fix ${\cal N}_a=1/s^a_1$. The entropy becomes
\begin{eqnarray}
    S_2(a) &=& 
    -\log \sum_b \frac{s^a_b}{\left(s^a_1\right)^2} 
                 {\rm Tr}~ \left(v_1^L\right)^T v_1^R \left(v_1^L\right)^T v_b^R.
\end{eqnarray} 
The trace is a tensor network in Fig. \ref{fig:entropyV} (A). It is equal to a trace of $L_v$-th power of a transfer matrix. For large enough $L_v$ the network becomes 
\begin{equation}
    {\rm Tr}~
    \left(v_1^L\right)^T v_1^R \left(v_1^L\right)^T v_b^R =~
    G_b~ \Lambda_b^{L_v}.
\end{equation} 
Here $\Lambda_b$ is the leading eigenvalue of the transfer matrix and $G_b$ its degeneracy. For large enough $L_v$ the entropy is dominated by terms with the maximal leading eigenvalue,
\begin{equation}
    \Lambda={\rm Max}_b~ \Lambda_b,
\end{equation}
and becomes
\begin{eqnarray}
    S_2(a) &=& 
    -\log 
    \Lambda^{L_v}
    \widetilde \sum_b
    \frac{G_bs^a_b}{\left(s^a_1\right)^2}
    \equiv
    \alpha L_v - \gamma_a.
\end{eqnarray}
Here the sum is restricted to indices $b$ with $\Lambda_b=\Lambda$. The area law has a coefficient
\begin{equation}
\alpha=-\log \Lambda
\end{equation}
that does not depend on anyon flux $a$ and the TEE is
\begin{equation}
    \gamma_a= 
    \log
    \widetilde \sum_b
    \frac{G_bs^a_b}{\left(s^a_1\right)^2}.
\end{equation}
We evaluate this expression in several examples.

\subsection{Toric code}

The projector yields $v_1^R\cdot P_\pm^T=(v_1^R\pm v_2^R)/2$, hence $s^\pm_1=1/2$ and $s^\pm_2=\pm 1/2$. Furthermore, we obtain
$
{\rm Tr}~
    \left( v_1^L \right)^T v_1^R \left( v_1^L \right)^T v_b^R = \Lambda^{L_v}    
$
when $b=1$ and zero otherwise. There is no degeneracy, $G_1=1$. Therefore,
\begin{equation}
    \gamma_\pm =\log \widetilde \sum_b 4 s^\pm_b = \log 4 s^\pm_1 = \log 2.
\end{equation}
This number is consistent with the anticipated identification $P_+\equiv P_{\rm vac}$ and $P_-\equiv P_e$.

\subsection{Fibonacci string net}

The projector yields $v_1^R\cdot P^T_\pm=(\phi^{\pm1} v_1^R\mp v_2^R)/\sqrt5$, hence $s^\pm_1=\phi^{\pm1}/\sqrt 5$ and $s^\pm_2=\mp 1/\sqrt 5$. We obtain with numerical precision:
\begin{eqnarray}
    \gamma_+ = \log {\cal D},~~
    \gamma_- = \log \frac{{\cal D}}{d_\tau d_{\bar\tau}},
\end{eqnarray} 
where ${\cal D}=2+\phi$ is the total quantum dimension and $d_\tau=d_{\bar\tau}=\phi$. These numbers are consistent with the identification $P_+\equiv P_{\rm vac}$ and $P_-\equiv P_{\tau\bar\tau}$.

\subsection{Ising string net}

Following similar lines for the double Fibonacci string net we obtain
\begin{eqnarray}
    \gamma_5 =\log {\cal D},~~~
    \gamma_6 =\log \frac{\cal D}{d_\psi d_{\bar\psi}},~~~
    \gamma_2 =\log \frac{\cal D}{d_\sigma d_{\bar\sigma}}  
\end{eqnarray}
with numerical precision. Here the total quantum dimension ${\cal D}=4$, $d_\sigma=d_{\bar\sigma}=\sqrt2$, and $d_\psi=d_{\bar\psi}=1$. They are consistent with the identifications: $P_5\equiv P_{\rm vac}$, $P_6\equiv P_{\psi\bar\psi}$, and $P_2\equiv P_{\sigma\bar\sigma}$.


\section{ Topological entropy: impurity projectors }
\label{sec:TEEimpurityprojectors}

For impurity projectors that act on an iPEPS that is inserted with $Z^h$ calculation of entropy goes along similar lines but with modifications accounting for $Z^h$. Accordingly, we begin with $\sigma^{L,R}=x^{L,R}_1$. Here $x^{L}_i$ and $x^{R}_j$ are MPO forms of impurity eigenstates $\left(x^{L}_i\right|$ and $\left|x^{R}_j\right)$, respectively. As usual, their left/right indices correspond to the bra/ket layer. The action of $\tilde P_a$ yields
\begin{equation}
    x^R_1 \cdot \tilde P_a^T = \sum_b \tilde s^a_b x^R_b.
\end{equation}
Here coefficients $\tilde s^a_b$ are real because $x^{R}_b$ are Hermitean. Taking into account normalization that follows from their biorthonormality, $\delta_{i_1i_2}=\left(x^L_{i_1}\right.\left|x^R_{i_2}\right)={\rm Tr}~\left( x^{L}_{i_1}\right)^T x^{R}_{i_2}$, the entropy in sector $a$ becomes
\begin{equation}
    S_2(a) = 
    -\log 
    \sum_b 
    \frac{ \tilde s^a_b }{\left( \tilde s^a_1 \right)^2}
    {\rm Tr}~
    \left( x_1^L \right)^T x_1^R \left( x_1^L \right)^T x_b^R .
\end{equation}
The trace is a trace of the tensor network in Fig. \ref{fig:entropyV} (B). It is a trace of $L_v$-th power of a transfer matrix times a layer of impurities $\bbX^{L,R}_b$. The transfer matrix is the same as in Fig. \ref{fig:entropyV} (A). For large enough cylinder width $L_v$ the sum is dominated by indices $b$ such that $\Lambda_b=\Lambda$, where $\Lambda$ is the same maximal leading eigenvalue of the transfer matrices:
\begin{equation}
    S_2(a) = \alpha L_v - \widetilde\gamma_a.
\end{equation}    
Here $\alpha=-\log\Lambda$ is the same as for vertical projectors and independent of anyon flux $a$. The topological entropy is
\begin{equation}    
    \widetilde \gamma_a=
    \log 
    \widetilde \sum_b
    \frac{ \tilde s^a_b }{\left( \tilde s^a_1 \right)^2}
    \sum_{m=1}^{G_b}
    X^a_{b,m}.
\end{equation}    
Here 
\begin{equation}    
    X^a_{b,m}=
    (U_{b,m}|
    {\rm Tr} \left( \bbX_1^L \right)^T \bbX_1^R \left( \bbX_1^L \right)^T \bbX_1^R 
    |D_{\{b_i\},m}), 
\end{equation}
is a form factor where $(U_{1,m}|$ and $|D_{1,m})$ are the up and down leading eigenvectors of the transfer matrix in Fig. \ref{fig:entropyV} (B), numbered by $m=1...G_b$ where $G_b$ is the degeneracy of the leading eigenvalue, and ${\rm Tr} \left( \bbX_1^L \right)^T \bbX_1^R \left( \bbX_1^L \right)^T \bbX_b^R$ is the MPO equal to the horizontal layer of impurities $\bbX_b^{L,R}$ in the same figure. The numerical procedure was applied in the following examples.

\subsection{Toric code}

The impurity projectors $\tilde P_\pm$ together with IMPO fusion rules (\ref{impurityZFR}) determine the coefficients $\tilde s_{\pm 1}=1/2$ and $\tilde s_{\pm 2}=\pm1/2$. As for vertical projectors, the truncated sum runs over $b=1$ only with degeneracy $G_1=1$. The topological entropies are
\begin{equation}    
    \widetilde\gamma_\pm = \log 2 X^a_{1,1} = \log 2,
\end{equation}   
within numerical precision. This number is obtained after numerical evaluation of the form factors and is consistent with the identification $\widetilde P_+=\widetilde P_{\rm m}$ and $\widetilde P_-=\widetilde P_{\rm \epsilon}$.

\subsection{Fibonacci string net}

Numerical evaluation of coefficients $\tilde s^a_b$ and the form factors yields
\begin{eqnarray}
    \widetilde \gamma_{1} &=& \log \frac{\cal D}{d_\tau} ,~~~
    \widetilde \gamma_{2} = \log \frac{\cal D}{d_{\bar\tau}} ,~~~
    \gamma_- = \log \frac{\cal D}{d_\tau d_{\bar\tau}}
\end{eqnarray} 
with numerical precision. Here ${\cal D}=2+\phi$ is the total quantum dimension and $d_\tau=d_{\bar\tau}=\phi$. These numbers are consistent with the identifications: $\widetilde P_{1}=\widetilde P_{\tau}$, $\widetilde P_{2}=\widetilde P_{{\bar\tau}}$, and $\widetilde P_3=\widetilde P_{\tau\bar\tau}$.

\subsection{Ising string net}

Similar numerical evaluation as for Fibonacci model yields 
\begin{eqnarray}
&&
    \widetilde \gamma_{\sigma} =          
        \log \frac{{\cal D}}{d_\sigma},~~
    \widetilde \gamma_{\bar\sigma} =  
        \log \frac{{\cal D}}{d_{\bar\sigma}},\\
&&        
    \widetilde \gamma_{\sigma\bar\psi} = 
        \log \frac{{\cal D}}{d_\sigma d_{\bar\psi}},~~
    \widetilde \gamma_{\psi\bar\sigma} = 
        \log \frac{{\cal D}}{d_\psi d_{\bar\sigma}},\\
&&        
    \widetilde \gamma_{\psi} = 
        \log \frac{{\cal D}}{d_\psi},~~
    \widetilde \gamma_{\bar\psi} = 
        \log \frac{{\cal D}}{d_{\bar\psi}},\\
&&        
    \widetilde \gamma_{\sigma\bar\sigma} =
        \log \frac{{\cal D}}{d_\sigma d_{\bar\sigma}}
\end{eqnarray}  
within numerical precision. Here the total quantum dimension is ${\cal D}=4$ while $d_\sigma=d_{\bar\sigma}=\sqrt2$ and $d_\psi=d_{\bar\psi}=1$. The numbers are consistent with the anticipated identification of the projectors.


\begin{figure}[t!]
\includegraphics[width=0.99\columnwidth]{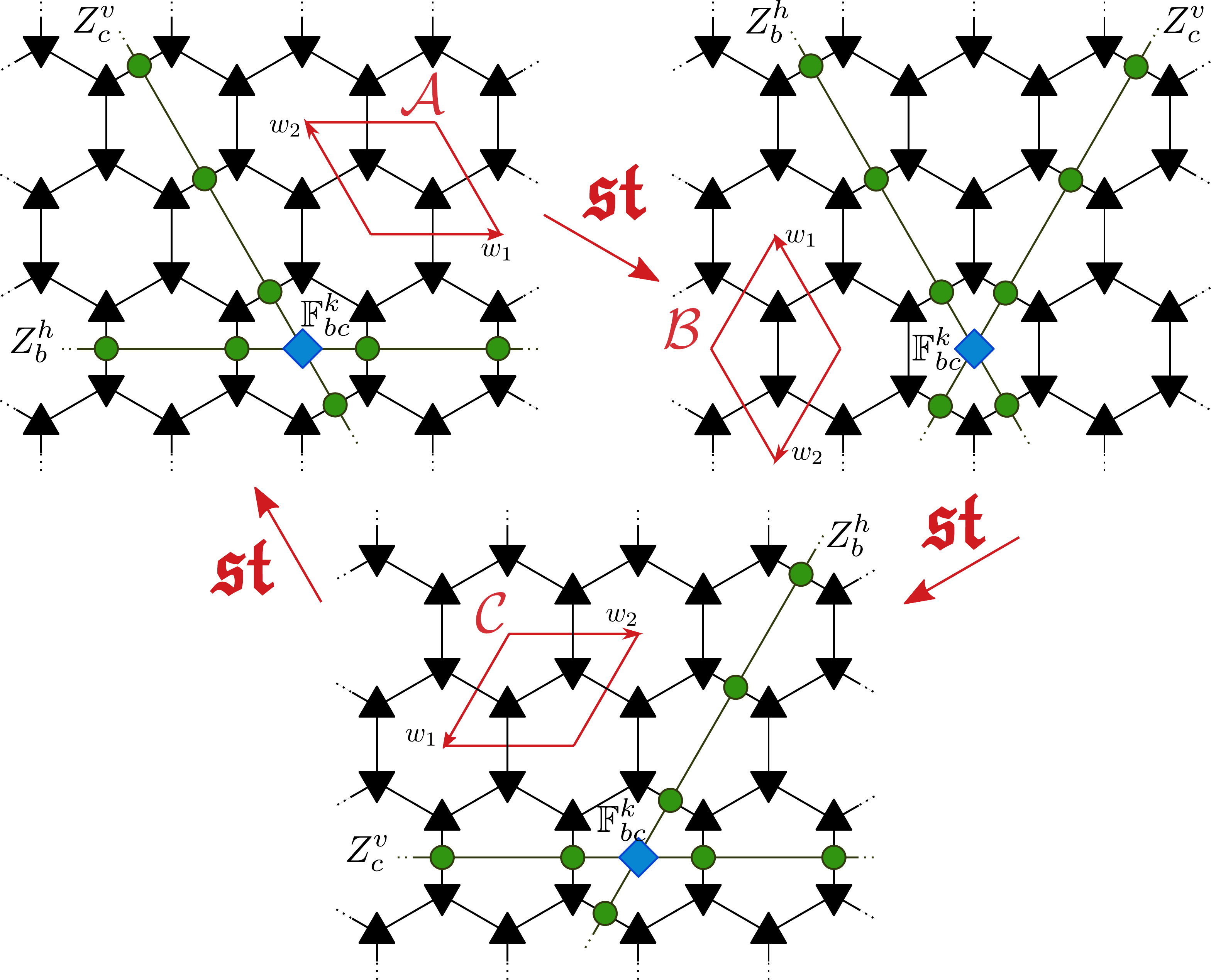}
\caption{
{\bf Basic state.}
The object
$
\bbF_{bc}^k,
$
includes the lines of $Z^h_b$ and $Z^v_c$ and a tensor at their intersection. 
When $b=1$ ($c=1$) then $\bbF$ is just vertical MPO symmetry $Z^v_c$ (horizontal $Z^h_b$). 
When $b>1$ then $\bbF^k_{bc}$ is one of the IMPO symmetries. 
Inserted into an iPEPS wrapped on an infinite torus it yields state $\ket{ \bbF_{bc}^k }$.
The same set of states (for each $b,c,k$) can be found on each of the tori related by modular $\mathfrak{st}$ transformation, where $(\mathfrak{st})^3=\mathbb{I}$, which corresponds to $120^\circ$ counterclockwise rotation on the honeycomb lattice with the chosen tori defined by a pair of unit vectors $(w_1,w_2)$.
}
\label{fig:X}
\end{figure}


\section{Topological $S$ and $T$ matrices}
\label{sec:STmatrices}

For pedagogical reasons, up to this point we distinguished between vertical projectors, with a trivial $Z^h_1=1^h$, and impurity projectors. For the present purpose of calculating topological $S$ and $T$ matrices it may be more convenient to treat them all on equal footing. We number MPO symmetries as $Z^{h,v}_a$ with $a=1,...,n$, where $a=1$ labels the trivial identities $1^{h,v}$. A basic building block for the projectors is 
$
\bbF_{bc}^k,
$
shown in Fig. \ref{fig:X}, including the lines of $Z^h_b$ and $Z^v_c$ and a tensor at their intersection. When $b=1$ ($c=1$) then $\bbF$ is just vertical MPO symmetry $Z^v_c$ (horizontal $Z^h_b$). 
When $b>1$ then $\bbF^k_{bc}$ is one of the IMPO symmetries. Therefore, in this unified notation each (vertical or impurity) projector on anyon flux $a$ can be expressed as a linear combination 
\begin{equation}
    P_a = \sum_{bc} \sum_{k} c^a_{kbc} \bbF_{bc}^k,
\end{equation}
where the range of $k$ depends on $bc$. When inserted into iPEPS wrapped on an infinite torus, the projector yields the ground state with anyon flux $a$ in the horizontal direction:
\begin{equation}
    \ket{ \Psi^a } = 
    \sum_{ab} \sum_k 
    c^a_{kab} 
    \ket{ \bbF_{ab}^k }.
\end{equation}
Here the last ket is the iPEPS inserted with $\bbF_{\alpha\beta}^k$. Up to this point there is nothing essentially new in this paragraph except for fixing notation.

States $\ket{\Psi^a}$ are used to calculate topological $S$ and $T$ matrices. Diagonal $T$ matrix encodes self-statistics, while $S$ matrix stands for mutual statistics. Together they form a representation of a modular group $SL(2,\mathbb{Z})$, by which they are related to the modular transformations of a torus generated by $\mathfrak{s}$ and $\mathfrak{t}$ transformations \citep{wen2015theory}. It follows that the matrix elements of a combination of the topological $S$ and $T$ matrices are given by the overlaps between $\ket{\Psi^a}$ transformed by a combination of corresponding modular matrices $\mathfrak{s}$ and $\mathfrak{t}$.

Here we work with states on a hexagonal lattice with $120^\circ$ rotational symmetry and we start by defining torus $\mA$ in Fig.~\ref{fig:X} with unit vectors $w_1$, $w_2$ and corresponding transfer matrices: vertical $(w_1,L_vw_2)$ and horizontal $(L_hw_1,w_2)$ with $L_{h,v} \rightarrow \infty$, see Fig.~\ref{fig:TM}(B) for comparison. Next, we consider all transformations of the unit cell by $\mathfrak{st}$ matrix, which generates $120^\circ$ counterclockwise rotation, see Fig.~\ref{fig:X}. This results in tori $\mB$ and $\mC$ together with their corresponding transfer matrices as shown in Fig.~\ref{fig:X}. This construction, however, is general and can be applied to lattices with other symmetries as well. 


Our method requires finding three complete sets of ground states 
\begin{equation} \label{eq:complete}
\left\{ \ket{\Psi^a_\mA} \right\}, \quad
\left\{ \ket{\Psi^a_\mB} \right\}, \quad
\left\{ \ket{\Psi^a_\mC} \right\}, 
\end{equation}
with well-defined anyon fluxes corresponding to three different tori: $\mA$, $\mB$, $\mC$. Topological $S$ and $T$ matrices are extracted from all possible overlaps between states in (\ref{eq:complete}). This algorithm is presented in \cite{zhang2015general} and slightly generalized in the appendix of Ref. \cite{topoAF1}. 


\begin{figure}[t!]
\includegraphics[width=0.99\columnwidth]{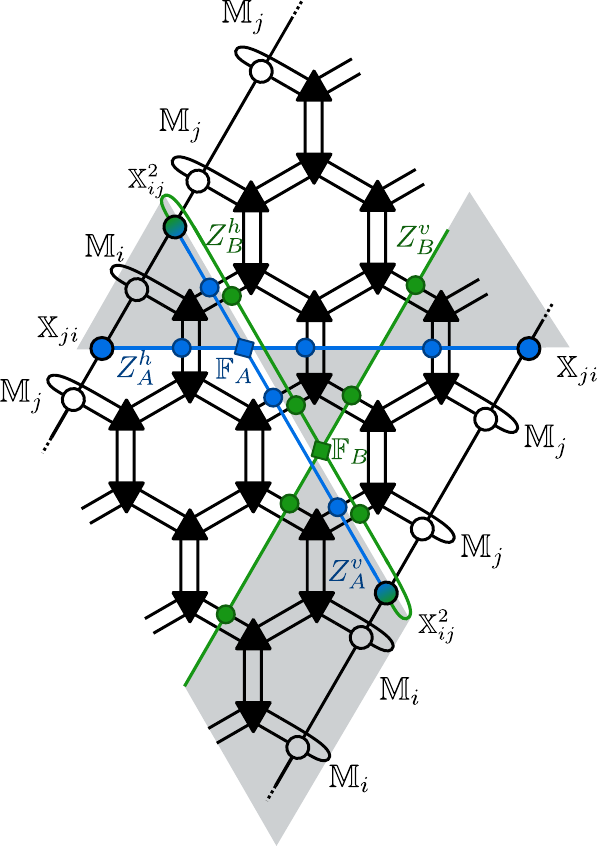}
\caption{
The overlap in Eq. (\ref{FAFB}) between iPEPS' on infinite tori $\mA$ and $\mB$ calculated on torus $\mB$, using its vertical boundary MPS. It involves new class of impurity transfer matrices and their eigenvectors, where a non-trivial MPO symmetry is in only one layer of the PEPS (either bra or ket) or there are two non-trivial MPO symmetries in both layers but they are of a different type. Inserting an MPO symmetry may in general change the boundaries, hence the change of indices $\mathbb{M}_i\rightarrow\mathbb{M}_j$ and the grey shaded regions denoting these sector changes.
}
\label{fig:FAFB}
\end{figure}


The core of the calculation is an overlap 
\begin{equation}
\left\langle 
\left( \bbF^k_{ab} \right)_\mA 
\right.\left|     
\left( \bbF^{k'}_{a'b'} \right)_\mB          
\right\rangle,
\label{FAFB}
\end{equation}
shown in Fig. \ref{fig:FAFB}, between two iPEPS's on infinite tori $\mA$ and $\mB$. It involves new class of impurity transfer matrices and their eigenvectors, where a non-trivial MPO symmetry is in only one layer of the PEPS (either bra or ket) or there are two non-trivial MPO symmetries in both layers but they are of a different type. This type of overlap was encountered already in the Abelian case in Ref. \cite{topoAF1} where they are explained in more detail. In Abelian case the non trivial MPO symmetry inserted in just one layer of the PEPS changes the boundary MPS $\vert v_i\rangle \rightarrow\vert v_j\rangle$, where $i\neq j$. However in the non Abelian case, all changes of the boundary MPS have to be considered including $i=j$. The possible change of the boundary conditions is denoted in the Fig.\ref{fig:FAFB} by shaded grey regions. Once the overlaps are found, we follow the algebra in appendix B of Ref. \cite{topoAF1} to obtain the following topological matrices $S$ and $T$.

\subsection{Toric code}

For analytic tensors with $D=4$ we obtain the exact matrices up to numerical precision:
\begin{equation*}
{\scriptsize S_{\rm{TC}} = \frac{1}{2}
\begin{pmatrix}
  1 & 1 & 1 & 1\\
  1 & 1 & -1 & -1\\
  1 & -1& 1& -1 \\
  1 & -1 & -1 & 1 \\
 \end{pmatrix},\qquad
 \scriptsize
  T_{\rm{TC}} = \begin{pmatrix}
  1 & 0 & 0 & 0 \\
  0 & 1 & 0 & 0 \\
  0 & 0 & 1 & 0 \\
  0 & 0 & 0 & -1 \\
 \end{pmatrix}} \ .
\end{equation*}
Here consecutive columns and rows correspond to projectors that were labelled as $1,e,m,\epsilon$.
This matrices confirm correctness of this labelling up to possible interchange of $e$ and $m$ that is a matter of convention.

\subsection{Fibonacci string net}

For the five states obtained with projectors $P_{\rm vac}$, $P_{\tau\bar\tau}$, $\widetilde P_{\tau\bar\tau}$, $\widetilde P_\tau$, $\widetilde P_{\bar\tau}$ we obtain the matrices:
\begin{eqnarray*}
S_{\rm{Fib}} &= \frac{1}{D}&
{\scriptsize\begin{pmatrix}
  1 &  \varphi^2 & \varphi^2 &  \varphi &  \varphi \\
  \varphi^2 &  1 & 1 & -\varphi & -\varphi \\
  \varphi^2 & 1 & 1 & -\varphi & -\varphi \\
  \varphi & -\varphi & -\varphi &  -1 & \varphi^2 \\
  \varphi & -\varphi & -\varphi &  \varphi^2 &  -1 
 \end{pmatrix}},\qquad \\
T_{\rm{Fib}} &=& 
 {\scriptsize\begin{pmatrix}
  1 & 0 & 0 & 0 & 0\\
  0 & 1 & 0 & 0 & 0\\
  0 & 0 & 1 & 0 & 0\\
  0 & 0 & 0 & \mathrm{e}^{4i\pi/5} & 0\\
  0 & 0 & 0 & 0 & \mathrm{e}^{-4i\pi/5}
 \end{pmatrix}} .
\end{eqnarray*}
For brevity matrix $S_{\rm Fib}$ is shown exact with $\varphi=d_\tau=\frac{1}{2}(1+\sqrt{5})$ although we obtain it with numerical accuracy $\mathcal{O}(10^{-10})$. It is clear that we can remove either second or third row and column because they both correspond to two equivalent ways of obtaining flux $\tau\bar\tau$.

\subsection{Ising string net}
For the ten states obtained with projectors $P_{\rm vac}$, $P_{\psi\bar\psi}$, $P_{\sigma\bar\sigma}$, $\widetilde P_{\sigma\bar\sigma}$, $\widetilde P_{\bar\psi}$, $\widetilde P_{\psi}$, $\widetilde P_{\sigma}$, $\widetilde P_{\sigma\bar\psi}$, $\widetilde P_{\bar\sigma}$, $\widetilde P_{\psi\bar\sigma}$ we obtain the matrices with numerical accuracy $\mathcal{O}(10^{-13})$:
\begin{eqnarray*}
S_{\rm{Is}} &= \frac{1}{4} &
{\scriptsize \begin{pmatrix}
    1 & 1 & 2 & 2 & 1 & 1 & \sqrt{2} & \sqrt{2} & \sqrt{2} & \sqrt{2}\\
    1 & 1 & 2 & 2 & 1 & 1 &-\sqrt{2} &-\sqrt{2} &-\sqrt{2} &-\sqrt{2}\\
    2 & 2 & 0 & 0 &-2 &-2 & 0 & 0 & 0 & 0\\
    2 & 2 & 0 & 0 &-2 &-2 & 0 & 0 &0 &0\\
    1 & 1 &-2 &-2 & 1 & 1 & \sqrt{2} & \sqrt{2} &-\sqrt{2} &-\sqrt{2}\\
    1 & 1 &-2 &-2 & 1 & 1 &-\sqrt{2} &-\sqrt{2} & \sqrt{2} & \sqrt{2}\\
    \sqrt{2} &-\sqrt{2} &0 &0 & \sqrt{2} &-\sqrt{2} & 0 & 0 & 2 & -2\\
    \sqrt{2} &-\sqrt{2} &0 &0 & \sqrt{2} &-\sqrt{2} & 0 & 0 & -2 & 2\\
    \sqrt{2} &-\sqrt{2} &0 & 0 &-\sqrt{2} & \sqrt{2} & 2 & -2 & 0 & 0\\
    \sqrt{2} &-\sqrt{2} &0 & 0 &-\sqrt{2} & \sqrt{2} & -2 & 2 & 0 & 0
    \end{pmatrix}},\qquad \\
T_{\rm{Is}} &=& 
 {\scriptsize \begin{pmatrix}
  1 & 0 & 0 & 0 & 0 & 0 & 0 & 0 & 0& 0\\
  0 & 1 & 0 & 0 & 0 & 0 & 0 & 0 & 0& 0\\
  0 & 0 & 1 & 0 & 0 & 0 & 0 & 0 & 0& 0\\
  0 & 0 & 0 & 1 & 0 & 0 & 0 & 0 & 0& 0\\
  0 & 0 & 0 & 0 & -1 & 0 & 0 & 0 & 0& 0\\
  0 & 0 & 0 & 0 & 0 & -1 & 0 & 0 & 0& 0\\
  0 & 0 & 0 & 0 & 0 & 0 &\mathrm{e}^{i\pi/8} & 0 & 0 & 0\\
  0 & 0 & 0 & 0 & 0 & 0 & 0 &-\mathrm{e}^{i\pi/8} & 0& 0\\ 
  0 & 0 & 0 & 0 & 0 & 0 & 0 & 0 & \mathrm{e}^{-i\pi/8} & 0 \\ 
  0 & 0 & 0 & 0 & 0 & 0 & 0 & 0 & 0 &-\mathrm{e}^{-i\pi/8} \\ 
 \end{pmatrix}} .
\end{eqnarray*}
It is clear that we can remove either third or fourth row and column because they both correspond to two equivalent ways of obtaining flux $\sigma\bar\sigma$. 


\section{Summary}
\label{sec:summary}

We presented numerical method to determine non-Abelian topological order in iPEPS representing the unique ground state on infinite two-dimensional lattice. The method is based on finding consecutively the following elements:

\begin{enumerate}

\item All of the boundary fixed points of PEPS transfer matrices in the form of matrix product operators $v_i$;

\item All MPO symmetries $Z_a$ mapping between the boundaries and their fusion rules;

\item All impurity eigenvectors $x_a$ of vertical impurity transfer matrices of PEPS inserted with horizontal MPO symmetries $Z^h$;

\item All impurity MPO symmetries $\widetilde Z$ mapping between the impurity eigenvectors;

\item All projectors on states with well defined anyon flux along horizontal direction. They are linear combinations of either vertical MPO symmetries or vertical impurity MPO symmetries: $P_a=\sum_{bc}\sum_k c^a_{kbc} \bbF^k_{bc}$;

\item All overlaps between states with definite anyon flux on different infinite tori related by modular transformations. 

\end{enumerate}

The topological charges and mutual statistics in the form of topological $S$ and $T$ matrices are recovered from the overlaps. They provide full topological characterization of string net models. 

A byproduct of the linear ansatz for a projector is an efficient algorithm to obtain the second Renyi topological entanglement entropy directly in the thermodynamic limit. In addition to tests for the string net models, we found non-zero TEE in the variational ansatz of Ref. \cite{chiralpeps} for the Kitaev model in magnetic field \cite{kitaev2006anyons}, see appendix \ref{sec:kitaev}. 


\acknowledgements 

We are indebted to Lukasz Cincio and Guifre Vidal, the coauthors of our common Ref. \cite{topoAF1}, for laying foundations for the present generalization. Special thanks to Lukasz for helpful comments on the present manuscript. We would also like to thank Hyun-Yong Lee for very useful feedback on the ansatz in Ref. \cite{chiralpeps}. AF would like to thank Bram Vanhecke for explaining the VUMPS algorithm.
Numerical calculations were performed in MATLAB with the help of \verb+ncon+ function \cite{NCON} for tensor contractions. 
AF acknowledges financial support by Polish Ministry of Science and Education, project No. DI2015 021345, from the budget funds for science in 2016-2020 under the Diamond Grant program.
This research was supported by Narodowe Centrum Nauki (NCN) under grant
2019/35/B/ST3/01028 
(AF, JD) and Etiuda grant 2020/36/T/ST3/00451 (AF). 

\bibliography{refs.bib}


\appendix




\section{Fusion rules}
\label{app:FR}
Fusion rules are encoded in $F$-symbols which have to satisfy the Pentagon equation:
\begin{figure}[h!]
\centering \includegraphics[scale=0.25]{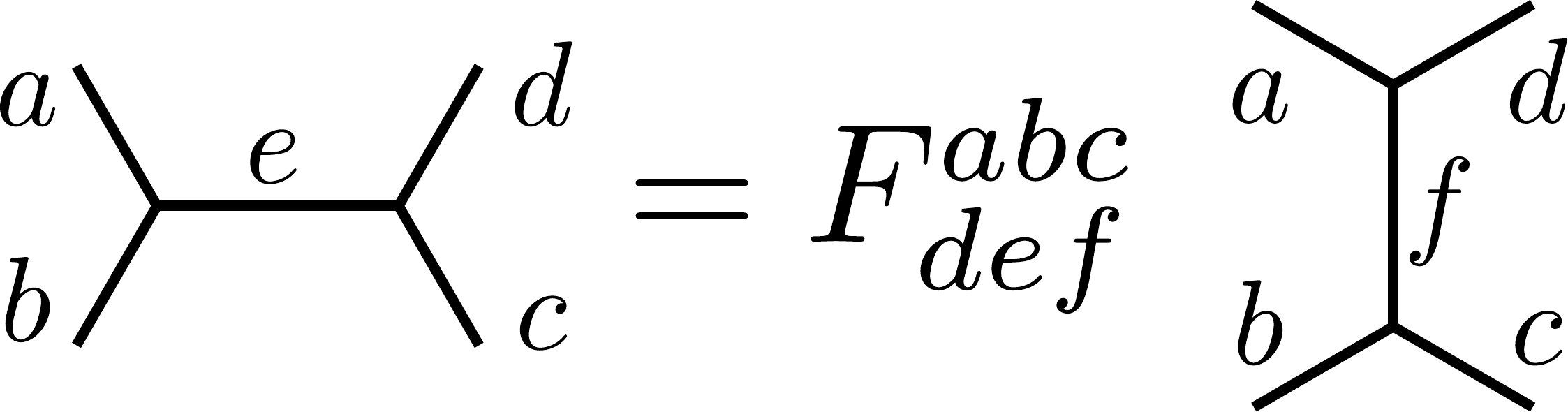}
\end{figure}\\
$F$-symbols of both non-Abelian models are mostly given by the allowed fusions: 
$N^c_{ab}$ describing $a\times b \rightarrow c$ with all its (allowed) permutations:
\begin{itemize}
\item for Fibonacci: $N^1_{11} = N^1_{\tau\tau} = N^\tau_{\tau\tau} = 1$
\item for Ising: $N^1_{11} = N^1_{\sigma\sigma} = N^1_{\psi\psi} = N^\sigma_{\psi\psi} = 1$.
\end{itemize}
Then $F^{abc}_{def} = N^e_{ab}N^e_{cd}N^f_{ad}N^f_{bc}$ unless they are overwritten by additional special rules:
\begin{itemize}
\item for Fibonacci: $F^{\tau\tau\tau}_{\tau11} = -F^{\tau\tau\tau}_{\tau\tau\tau} = \frac{1}{d_\tau}$ and $F^{\tau\tau\tau}_{\tau\tau1} = F^{\tau\tau\tau}_{\tau1\tau} = \frac{1}{\sqrt{d_\tau}}$.
\item for Ising: $F^{\sigma\sigma\sigma}_{\sigma11} = F^{\sigma\sigma\sigma}_{\sigma1\psi} = F^{\sigma\sigma\sigma}_{\sigma\psi1} = -F^{\sigma\sigma\sigma}_{\sigma\psi\psi} = \frac{1}{\sqrt{2}}$ and $F^{\psi\sigma\psi}_{\sigma\sigma\sigma} = F^{\sigma\psi\sigma}_{\psi\sigma\sigma} = -1$.
\end{itemize}

\begin{figure}[t!]
\centering
\includegraphics[width=0.85\columnwidth]{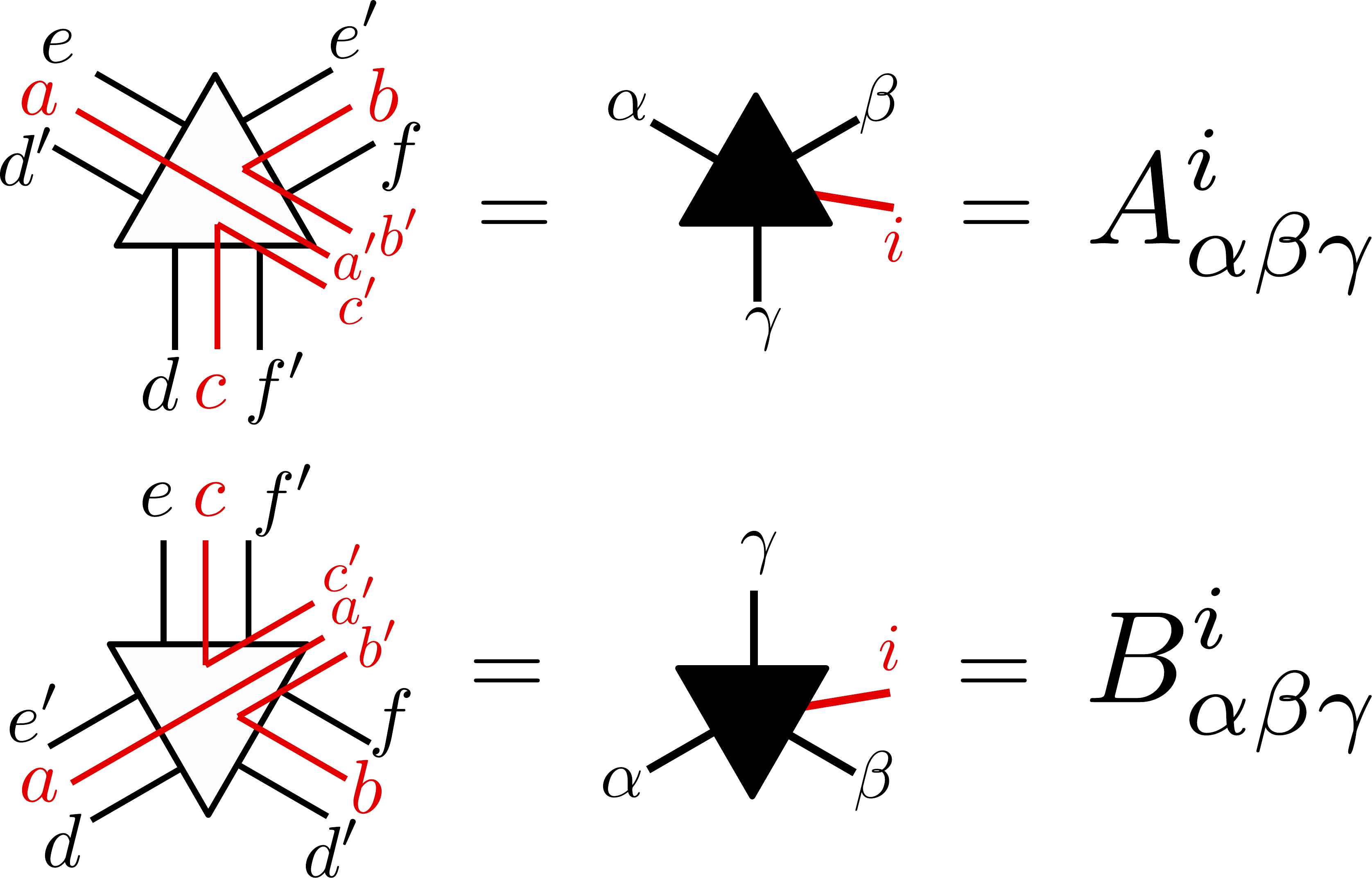}
\caption{
Tensors forming the iPEPS are defined via combination of $F$-symbols and corresponding quantum dimensions $d_i$. All bond indices and the physical index are in fact a triple index. The bond dimension can be reduced by applying projectors on the non zero bond indices.
}
\label{fig:tens}
\end{figure}

\section{iPEPS tensors}
\label{app:tensors}
iPEPS tensors, shown in Fig. \ref{fig:tens} are given by the following combination of $F$-symbols and quantum dimensions $d_i$:
\begin{eqnarray}
A^i_{\alpha\beta\gamma} &=& \left(\frac{d_ad_b}{d_c}\right)^{1/4}F^{dab}_{fec}\delta_{aa'}\delta_{bb'}\delta_{cc'}\delta_{dd'}\delta_{ee'}\delta_{ff'} ~~\\
B^i_{\alpha\beta\gamma}  &=& \left(\frac{d_ad_b}{d_c}\right)^{1/4}F^{dab}_{fec}\delta_{aa'}\delta_{bb'}\delta_{cc'}\delta_{dd'}\delta_{ee'}\delta_{ff'} ~~
\end{eqnarray}
By construction each tensor has a triple of bond indices along each of the three bonds towards NN lattice sites. We concatenate each triple into a single bond index, e.g., $\alpha=(a,e,d')$. The physical index is also a triple index $i=(a',b',c')$.
These basic tensors are forming the topological state after proper contraction of bond indices with respect to their triplet structure. For the toric code and double Fibonacci string nets the bond dimension $D=2^3=8$ is redundantly large and can be reduced to $D=4$ and $D=5$ after applying  projectors on the bond indices, namely the only non-zero combinations of bond indices $(i,j,k)$ are those, in which the fusion product $i\times j\times k = 1 +...$ contains the trivial anyon. For the double Ising string net, on the other hand, the original bond dimension $D=3^3=27$ can be reduced to $D=10$.

\begin{figure}[t!]
    \centering
    \includegraphics[width=0.85\columnwidth]{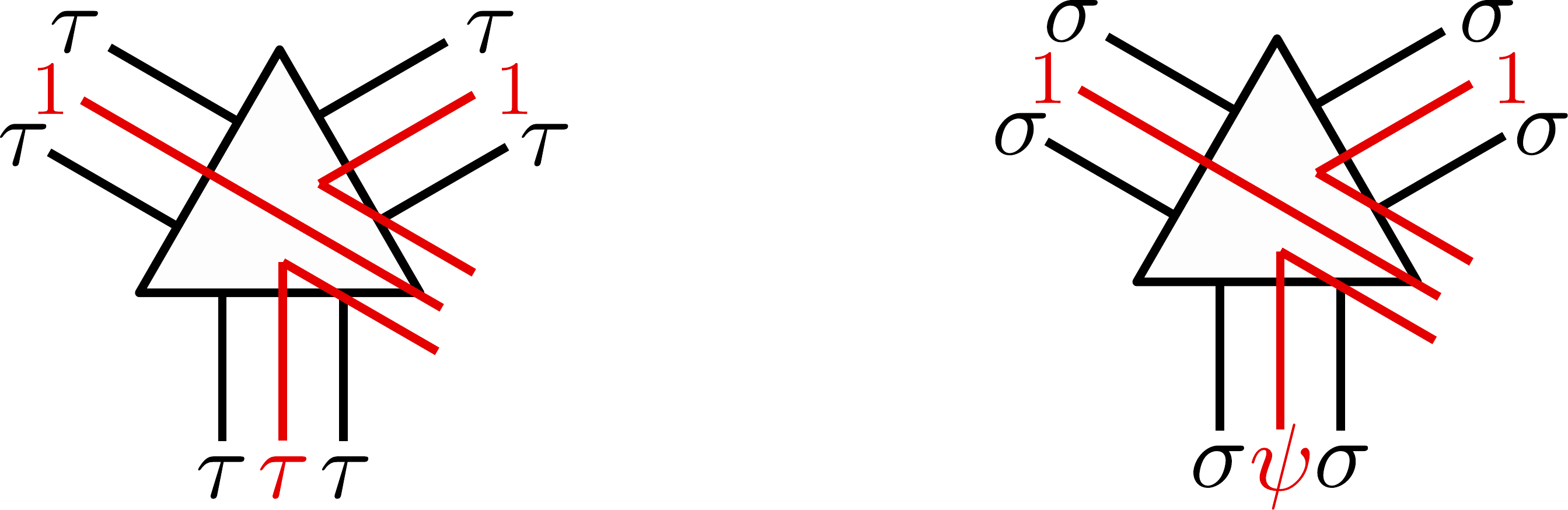}
    \caption{Symmetry breaking perturbations in Fibonacci (left) and Ising (right) string nets.}
    \label{fig:symm_breaking}
\end{figure}

\section{Perturbation of tensor symmetry}
\label{sec:symm_breaking}

In Ref. \onlinecite{topoAF1} we demonstrated that with our method it is possible to obtain accurate results for topological $S$ and $T$ matrices from a numerically optimized iPEPS ground state of the Kitaev honeycomb model for a wide range of coupling parameters. In case of Fibonacci and Ising string-nets, whose parent Hamiltonians are far more complex, the same test would go far beyond the scope of the present paper. However, as most concerns about stability arise from Refs. \onlinecite{symmetry_perturbation}, we can introduce their perturbation at the virtual level of the tensor network --- which violates the exact MPO-symmetries --- to see how our algorithm performs under this crash test.

The vertex violating terms \cite{symmetry_perturbation}, $T_p$, which are allowed in the stand-alone space but do not represent the physical ground state, are shown in Fig. \ref{fig:symm_breaking}. Additionally we allow all three rotations of the red indices. The fixed-point tensors $T$ are perturbed by adding a vertex violating term $T_p$ controlled by a small parameter $\epsilon$:
\begin{equation}
    T ~\rightarrow~ T+\epsilon~ T_p.
\end{equation}
For the Fibonacci string-net model perturbed with a strong $\epsilon = 0.1$ we obtained the following topological entanglement entropies:
\begin{eqnarray}
  \begin{pmatrix}
  1.2847 \\
  0.3235 \\
  0.3235 \\
  0.8047 \\
  0.8047 \\
  \end{pmatrix},
\end{eqnarray}
and the following topological matrices: 
\begin{equation*}
S_{\rm Fib}=
    \begin{pmatrix}
    0.2771  &  0.7251 &   0.7252  &  0.4484 &   0.4484 \\
    0.7251  &  0.2735 &   0.2749  & -0.4486 &  -0.4486  \\
    0.7252  &  0.2749 &   0.2764  & -0.4472 &  -0.4472 \\
    0.4484  & -0.4486 &  -0.4472  & -0.2764 &   0.7236 \\
    0.4484  & -0.4486 &  -0.4472  &  0.7236 &  -0.2764 \\
    \end{pmatrix}
\end{equation*}
and
\begin{equation*}
{\rm diag}\left( T_{\rm Fib} \right)=
    \begin{pmatrix}
   1.0000 - 0.0000i \\
   1.0000 + 0.0000i \\
   1.0000 - 0.0000i \\
  -0.8090 - 0.5878i\\
  -0.8090 + 0.5878i\\
    \end{pmatrix}.
\end{equation*}
When compared to the exact numbers, their maximal error is of the order of $10^{-3}$. Although there are $4$ anyon fluxes in the Fibonacci model, here as in the main text we keep both $\mathcal{P}_{\tau\bar\tau}$ and $\tilde{\mathcal{P}}_{\tau\bar\tau}$ which project on the same flux $\tau\bar\tau$. 

For the Ising string-net model we added a perturbation shown in Fig.\ref{fig:symm_breaking} with strength $\epsilon =0.5$, which lead to even more accurate results. We obtained topological entanglement entropy  and topological $S$ and $T$ matrices with accuracy $\mathcal{O}(10^{-6})$.

In order to complete the discussion about random perturbations that may arise during numerical optimization of iPEPS we calculated the topological data for a completely random, real perturbation in the Fibonacci string-net model:
\begin{equation}
    T ~\rightarrow~ T+\epsilon~ T_{random}.
\end{equation}
For $\epsilon = 0.01$ we recovered the topological entanglement entropies and topological matrices with accuracy of the order of $\mathcal{O}(10^{-4})$.

\begin{figure}[t!]
\centering
\includegraphics[width=0.90\columnwidth]{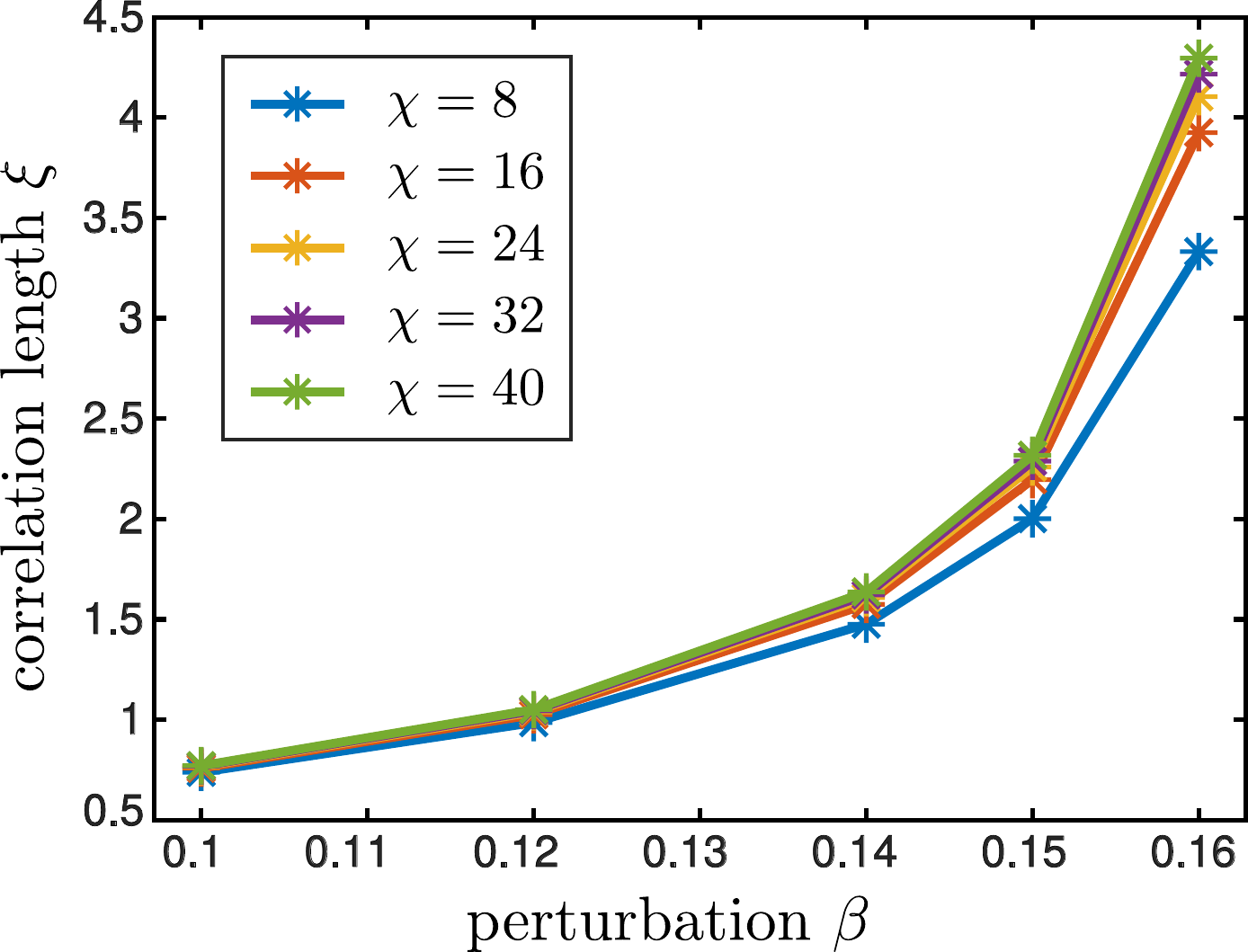}
\caption{
The correlation length $\xi$ in the Fibonacci string net model in function of the perturbation parameter $\beta$ in Eq. (\ref{beta}). Different colors correspond to different bond dimensions $\chi$ of the boundary MPS $v_i$.
}\label{fig:xi}
\end{figure}

\begin{table}[t!]
\begin{tabular}{|c|c|c|c|c|}
\hline
$\beta$ & $\xi$  & $\epsilon_\gamma$       & $\epsilon_S$            &       $\epsilon_T$      \\ \hline
0       &   0    & $\mathcal{O}(10^{-10})$ & $\mathcal{O}(10^{-10})$ & $\mathcal{O}(10^{-10})$ \\ \hline
0.14    &   1.64 & $\mathcal{O}(10^{-3 })$ & $\mathcal{O}(10^{-4})$  & $\mathcal{O}(10^{-6})$ \\ \hline
0.15    &   2.32 & $\mathcal{O}(10^{-2 })$ & $\mathcal{O}(10^{-3})$  & $\mathcal{O}(10^{-7})$ \\ \hline
0.16    &   4.3  & $\mathcal{O}(10^{-2 })$ & $\mathcal{O}(10^{-3})$  & $\mathcal{O}(10^{-4})$ \\ \hline
\end{tabular}
\caption{
For different values of the perturbation parameter $\beta$ in Eq. (\ref{beta}), 
the table lists corresponding correlation lengths, $\xi$, 
and maximal errors of the entries of the list of topological entanglement entropies, $\epsilon_\gamma$, 
and the $S$ and $T$ matrices, $\epsilon_S$ and $\epsilon_T$. 
}
\label{tab:beta}
\end{table}

\section{Introducing finite correlation length}
\label{sec:symm_breaking}

In order to see how the algorithm performs when the iPEPS tensors are driven away from the fixed point by introducing a finite correlation length, we apply the local filtering introduced in Refs. \onlinecite{per1,per2,per3} to the fixed point of the Fibonacci string-net model. The perturbation has the following form:
\begin{equation}
    \vert \Psi\rangle ~\rightarrow~ \prod_i \mathrm{e}^{\beta\sigma^z_i}\vert \Psi\rangle, 
    \label{beta}
\end{equation}
where the index $i$ runs over all physical indices and $\sigma^z$ is the third Pauli matrix.
In Ref. \onlinecite{topoAF1}, by considering a similar perturbation to the toric code, we demonstrated that with our algorithm it is possible to obtain topological $S$ and $T$ matrices for states with correlation length much longer than achievable by the state of the art 2D DMRG techniques. 

Figure \ref{fig:xi} shows how the correlation length grows with parameter $\beta$ for the perturbed Fibonacci string-net model. In the Fibonacci model, for parameters $\beta = 0.14, 0.15, 0.16$ such that the correlation length $\xi > 1$, we obtained the topological entanglement entropies and the topological $S$ and $T$ matrices. Their maximal errors are listed in table \ref{tab:beta}. 

\section{Variational ansatz for the Kitaev model in $(1,1,1)$ magnetic field} 
\label{sec:kitaev}

We investigate the ansatz proposed in the supplementary material of Ref. \cite{chiralpeps}. Although it satisfies all desired symmetries and has competitive energy, the ansatz was not demonstrated to possess the expected chiral Ising universality class \cite{kitaev2006anyons}. We show that at least it has non-trivial topological entanglement entropy. 

Each TM has two boundary fixed points. They have large bond dimension $\chi$ necessary to accommodate a long correlation length. For $\chi=150$ the correlation length saturates at $\xi\simeq15.4$. However, when it comes to calculating the topological entanglement entropy, whose cost is much steeper in $\chi$, we  will be satisfied with $\chi=50$, corresponding to $\xi\simeq 10.3$, that is sufficient to recover exact symmetries. There is one non-trivial ${\cal Z}_2$ symmetry such that $v_1^L\cdot Z_2^v=v_2^L$ and $v_2^L\cdot Z_2^v=v_1^L$ and, consequently, 
\begin{equation} 
Z_2^v\cdot Z_2^v=1^v.
\label{chiralFR}
\end{equation}
This is the algebra of the ${\cal Z}_2$ gauge field that was implemented in the ansatz by construction.

Like in the toric code, the ${\cal Z}_2$ algebra (\ref{chiralFR}) allows for two vertical projectors:
\begin{equation}
    P_{\pm}=\frac12\left(1^v\pm Z^v\right).
\end{equation}
They project on $\pm1$ horizontal flux of the ${\cal Z}_2$ gauge field, see Ref. \cite{chiralpeps}. In this model, when the horizontal cylinder is closed into a torus, the vertical flux also becomes a good quantum number. For an iPEPS wrapped on a torus (without horizontal line $Z^h_2$) the state is a superposition of both $\pm1$ vertical fluxes with equal amplitudes.   

We also find nontrivial IMPO symmetry $\widetilde Z^v_2$ satisfying the ${\cal Z}_2$ algebra. It allows for two projectors:
\begin{equation}
    \widetilde P_{\pm}=\frac12\left(\widetilde 1^v\pm \widetilde Z^v_2\right).
\end{equation}
Like the vertical projectors, they project on $\pm1$ horizontal flux of the ${\cal Z}_2$ gauge field, but with a superposition of vertical fluxes with opposite amplitudes. Therefore, unlike the Fibonacci and Ising string net, neither of these two impurity projectors can be identified with any of the two vertical projectors $P_\pm$.

For vertical projectors we obtain topological entanglement entropy
\begin{equation}
    \gamma_\pm =\log2
    \label{gammaKit}
\end{equation}
in the vacuum and vortex sector, respectively. This demonstrates topological order in the variational iPEPS of Ref. \cite{chiralpeps}. The impurity projectors also yield
\begin{equation}
    \tilde \gamma_\pm = \log 2
\end{equation}
but here the minimally entangled states $\pm$ are different combinations of the vertical ${\cal Z}_2$ flux than in Eq. (\ref{gammaKit}).

\end{document}